\documentclass[aps,twocolumn]{revtex4-1}
\usepackage{amsmath,amsthm,amssymb,amsfonts}
\usepackage{latexsym}
\usepackage{graphicx,psfrag,import}
\usepackage{fullpage}
\usepackage{framed}
\usepackage{verbatim}
\usepackage{color}
\usepackage{epsfig}
\usepackage{hyperref}
\usepackage{geometry}
\usepackage{mathtools}
\usepackage{float}
\usepackage{enumerate}
\usepackage{enumitem}
\usepackage{epstopdf}
\usepackage{mathrsfs}
\usepackage{subfigure}
\usepackage{caption}

\newcommand{\ket}[1]{| #1 \rangle}

\theoremstyle{plain}
\newtheorem{theorem}{Theorem}

\newtheorem{corollary}[theorem]{Corollary}

\theoremstyle{definition}

\date{}

\begin{document}

\title{\textbf{Measurement Induced Randomness and State Merging}}

\author{Indranil Chakrabarty}
\affiliation{Center for Security, Theory and Algorithmic Research, International Institute of Information Technology Hyderabad, Gachibowli, Hyderabad, Telangana 500032, India}
\author{Abhishek Deshpande}
\affiliation{Center for Computational Natural Sciences and Bioinformatics, International Institute of Information Technology Hyderabad, Gachibowli, Hyderabad, Telangana 500032, India}
\author{Sourav Chatterjee}
\affiliation{Center for Computational Natural Sciences and Bioinformatics, International Institute of Information Technology-Hyderabad, India} 
\affiliation{SAOT, Erlangen Graduate School in Advanced Optical Technologies, Paul-Gordan-Strasse 6, 91052 Erlangen, Germany}

\begin{abstract}

In this work we introduce the randomness which is truly quantum mechanical in nature arising as an act of measurement. For a composite classical system we have the joint entropy to quantify the randomness present in the total system and that happens to be equal to the sum of the entropy of one subsystem and the conditional entropy of the other subsystem given we know the first system. 
 The same analogy caries over to the quantum setting by replacing the Shannon entropy by the Von Neumann entropy. However, if we replace the conditional von Neumann entropy by the average conditional entropy due to measurement, we find that it is different from the joint entropy of the system. We call this difference “Measurement Induced Randomness (MIR)” and argue that this is unique of quantum mechanical systems and there is no classical counterpart to this. In other words the joint Von Neumann entropy gives only the total randomness that arises because of the heterogeneity of the mixture and we show that it is not the total randomness that can be generated in the composite system. We generalize this quantity for N-qubit systems and show that it reduces to quantum discord for two qubit systems. Further, we show that it is exactly equal to the change in the cost of quantum state merging that arises because of the  measurement. We argue that for a quantum information processing tasks like state merging the change in the cost as a result of discarding prior information can also be viewed as a rise of randomness due to measurement.

\end{abstract}
\maketitle

\section{Introduction}

Quantum entanglement~\cite{EPR,BELL,horodecki2009quantum} plays a crucial role as a resource in various information processing tasks like quantum teleportation~\cite{CHB}, cryptography~\cite{NG}, superdense coding~\cite{CHB2}, remote state preparation~\cite{AKP}, broadcasting of entanglement~\cite{IND1} and quantum state merging~\cite{HOW}. Subsequently researchers came up with different entanglement monotones like negativity~\cite{VID} and concurrence~\cite{SH,WW} as measures to quantify the amount of entanglement present in the system. However, there exist open issues in comprehending the nature of quantum correlations present in certain mixed and multi-qubit states. Information processing tasks can be performed even in the near absence of entanglement~\cite{KNILL}. As a result, innovative measures like quantum discord and dissension were introduced to elucidate the nature of quantum correlations~\cite{OLI,REST,DAK,ADHI,NAPO}.  

One important information processing task is state merging~\cite{HOW}. The idea of state merging is essentially the following: Consider two parties Alice and Bob. Bob has some prior information $Y$ and Alice has some missing information $X$ (where $X$ and $Y$ are random variables). If Bob wants to learn about $X$, how much additional information does Alice need to send him? It has been shown that only $H\left(X\mid Y\right)$ bits suffices. The same idea transfers over to the quantum setting by replacing the Shannon entropy by the Von neumann entropy. In the quantum setting, Alice and Bob each possess a system in some unknown quantum state with joint density operator $\rho_{AB}$. One asks how much additional quantum information Alice needs to send him, so that he knows the entire state. The amount of partial quantum information that Alice needs to send Bob is given by the quantum conditional entropy $S\left(A\mid B\right) = S\left(A,B\right) - S\left(B\right)$.

In classical systems, we use Shannon entropy to quantify the randomness associated with the physical system. For quantum mechanical systems, the Shannon entropy gets replaced by the Von neumann entropy. In the current manuscript, we show that in addition to the randomness associated with the heterogeneity of the mixture, there is true quantum mechanicalrandomness which arises due to the act of measurement. We call this as ``Measurement Induced Randomness"(MIR). We provide a way to quantify this randomness by replacing the quantum conditional entropy by an average entropy - averaged over projective measurments. In addition, 
we show that measurement 
induced randomness is responsible for the increase in the cost of state merging as a result of discarding prior information. 
In other words the cost of losing prior information in the state merging is equivalent of rise of randomness 
as a result of measurement. 

In Section~\ref{MIU}, we introduce the notion of Measurement Induced Randomness for two qubit, three qubit and finally for $N$ qubit systems. In Section~\ref{state_merging} we show that how measurement induced randomness is connected with state merging. Finally we conclude in Section~\ref{conclusion}.

\section{Measurement Induced Randomness}\label{MIU}

The entropy is an useful indicator of the amount of randomness present in the system. In  classical information theory, the randomness is quantified by Shannon entropy. In quantum mechanics we measure the amount of randomness generated as a result of representation of non homogeneous 
ensemble of states by the Von Neumann entropy of the state. For a composite classical system there is joint entropy to 
capture the total randomness present in the  system. Mathematically, this is equal to the sum of the entropy of one subsystem 
and the conditional entropy of the other subsystem provided we have the knowledge about the first system. In quantum mechanics, 
the only way to know about the state is to do a measurement on the state. Ideally the quantum analogy of conditional entropy 
will not work if we want to capture the randomness of the system provided we have knowledge about the system. Moreover once we 
do a measurement on the state, the state changes.  Here we replace the conditional entropy by a new quantity called the average 
entropy to quantify the randomness in one system with the notion of measurement already done in the other system.
Interestingly 
we find that if we replace conditional entropy by average entropy the total randomness in the system is not going to be equal to the 
joint entropy of the system. We call this difference as a randomness generated as a result of measurement (MIR). 


\subsection{Two Qubit}

For a composite system $\left(A_1,A_2\right)$, the classical randomness present in the system is given by the joint entropy $U_1 = H\left(A_1,A_2\right)$. Equivalently we can interpret the total randomness of $\left(A_1,A_2\right)$ as randomness of $A_1: H(A_1)$ and the randomness of $A_2$ given that we know about $A_1:H\left(A_2\mid A_1\right)$. The sum of these two quantities given by $U_2=H\left(A_1\right) + H\left(A_2\mid A_1\right)$ (In principle $U_2$ can also have an equivalent expression $U_2=H\left(A_2\right) + H\left(A_1\mid A_2\right)$. Classically, the expressions $U_1$ and $U_2$ are identical. However, if we replace the random variables $A_1$ and $A_2$ by quantum states $\rho_{A_1}$ and $\rho_{A_2}$ and Shannon entropies  $H(.)$, by Von Neumann entropies $S(.)$, these two expressions are no longer the same. In the quantum case involving measurements, the conditional entropy $H(A_2\mid A_1)$ is replaced by the entropy $S(\rho_{A'_2\mid A'_1})$, which is the average entropy obtained after carrying out a projective measurement on subsystem $A_1$. 
The projective measurement is done in the general basis $\{|u_1\rangle =\cos\theta|0\rangle + e^{i\phi}\sin\theta|1\rangle$, $|u_2\rangle =\sin\theta|0\rangle - e^{i\phi}\cos\theta|1\rangle\}$, where $\theta,\phi\in\left[0,2\pi\right]$. Hence, the average conditional entropy can be expressed as $S\left(\rho_{A'_2\mid A'_1}\right)=\displaystyle\sum_jp_jS\left(\rho_{A_2\mid\pi_j^{A_1}}\right)$ where $p_j=\text{tr}[(I_{A_2}\otimes\pi_j^{A_1})\rho(I_{A_2}\otimes\pi_j^{A_1})]$ and $\rho_{A_2\mid\pi_{j}^{A_1}}=\frac{1}{p_j}\text{tr}_{A_1}[(I_{A_2}\otimes\pi_j^{A_1})\rho(I_{A_2}\otimes\pi_j^{A_1})]$. The quantum mechanical expressions for $U_1$ and $U_2$ for two qubit system are given by, 

\begin{eqnarray*}
\begin{split}
U_1 &= S(\rho_{A_1A_2}),\\
U_2 &= S(\rho_{A_1}) + S(\rho_{A'_2\mid A'_1}).
\end{split}
\end{eqnarray*}

The difference between these two quantities is the measurement induced randomness (MIR):

\begin{eqnarray}\label{first_MIU}
\begin{split}
\Delta_{R}^{A_1A_2} &= U_2-U_1 \\
                    &= S(\rho_{A_1}) + S(\rho_{A'_2\mid A'_1}) - S(\rho_{A_1A_2}).  
\end{split}
\end{eqnarray}

On interchanging the qubits $A_1$ and $A_2$ we have a different expression for MIR as
\begin{eqnarray}\label{second_MIU}
\Delta_{R}^{A_1A_2}=S(\rho_{A_2}) + S(\rho_{A'_1\mid A'_2}) - S(\rho_{A_1A_2})
\end{eqnarray}

We show that for two qubit systems the minimum value of MIR is exactly equal to the discord. The two different discords~\cite{OLI} are given by, 

\begin{eqnarray*}
D_1= S(\rho_{A_1})-S(\rho_{A_1A_2})+\min_{\pi_j}S(\rho_{A'_2\mid A'_1}) \\
D_2= S(\rho_{A_2})-S(\rho_{A_1A_2})+\min_{\pi_j}S(\rho_{A'_1\mid A'_2})
\end{eqnarray*}

Using Equation~\ref{first_MIU}, we get $\displaystyle\min_{\pi_j}\Delta_{R}^{A_1A_2}=D_1$ and using Equation~\ref{second_MIU}, we have $\displaystyle\min_{\pi_j}\Delta_{R}^{A_1A_2}=D_2$. It is interesting to note both these discords are a positive quantities. 


\subsection{Three Qubit}

In this subsection we extend the idea of the measurement induced randomness (MIR) for three qubit systems. One possible expression for $U_2$ is
\begin{eqnarray*}
U_2=  S(\rho_{A_1})+S(\rho_{A'_2\mid A'_1})+S(\rho_{A'_3\mid A'_1A'_2}).
\end{eqnarray*}
This quantifies the total amount of randomness present in three qubit system given by the sum of the randomness in the system $A_1$, the randomness in the system $A_2$ once we know about the system $A_1$ and after that the randomness in the system $A_3$ given that we know about the system $A_1,A_2$. Here $S(\rho_{A'_2\mid A'_1})$ is the average entropy obtained by carrying out projective measurement on the subsystem $A_1$ and $S(\rho_{A'_3 \mid A'_1,A'_2})$ is the average entropy of the qubit $A_3$ obtained after carrying out a two particle projective measurement on the subsystem $\left(A_1,A_2\right)$. In principle, one can have different expressions for the total entropy depending upon the choice and sequence of measurements. The two-particle projective measurement is performed in the general two qubit basis: 
\begin{eqnarray*}
\mid v_1\rangle = \cos\theta\mid 00\rangle + \sin\theta\mid 11\rangle, \\
\mid v_2\rangle = \sin\theta\mid 00\rangle - \cos\theta\mid 11\rangle, \\
\mid v_3\rangle = \cos\theta\mid 01\rangle + \sin\theta\mid 10\rangle, \\
\mid v_4\rangle = \sin\theta\mid 01\rangle - \cos\theta\mid 10\rangle,
\end{eqnarray*}
where $\theta\in\left[0,2\pi\right]$. In this case, $S(\rho_{A'_3\mid A'_1,A'_2}) = \displaystyle\sum_jp_jS(\rho_{A_3\mid\pi_j^{A_1A_2}})$. Alternatively the expression for $U_1$ in a three qubit system is given by $U_1= S(\rho_{A_1,A_2,A_3})$. The MIR for three qubit systems is given by,
\begin{eqnarray*}
\begin{split}
\Delta_{R}^{A_1A_2A_3} &= U_2-U_1 \\
                       &= S(\rho_{A_1})+ S(\rho_{A'_2\mid A'_1}) +\\ &S(\rho_{A'_3\mid A'_1A'_2}) - S(\rho_{A_1A_2A_3}).
\end{split}      
\end{eqnarray*}


\begin{corollary}\label{lemma2}
\noindent Let $\rho_{A_1A_2A_3}$ be a pure three qubit state. Then the measurement induced randomness associated with this state is given by $\Delta_{R}^{A_1A_2A_3} = S(\rho_{A_1})+ S(\rho_{A'_2\mid A'_1})$.
\end{corollary}

\begin{proof}
For arbitrary pure three qubit state $\rho_{A_1A_2A_3}$, $S(\rho_{A_3\mid A_1A_2})=0$. This is because after measurement the system is in a product state of the state $A_3$ and the projected state $(A_1,A_2)$, which is a pure state. In addition, we have $S(\rho_{A_1A_2A_3})=0$ since $\rho_{A_1A_2A_3}$ is a pure state. Therefore, we have

\begin{eqnarray*}
\begin{split}
\Delta_{R}^{A_1A_2A_3} &= S(\rho_{A'_3\mid A'_1A'_2})- S(\rho_{A_1A_2A_3})+\\
&S(\rho_{A_1})+S(\rho_{A'_2\mid A'_1}) \\
					   &= S(\rho_{A_1})+S(\rho_{A'_2\mid A'_1}).
\end{split}
\end{eqnarray*}

\end{proof}

\subsection{N qubit}

Finally we introduce the concept of measurement induced randomness for $N$ qubit systems. As observed in previous cases, we also have different expressions of $U_2$ depending on the choice of qubit on which the measurement is performed. The total amount of randomness in the composite system when no measurement is performed is given by $U_1=S(\rho_{A_1A_2...A_N})$. Now if we want to quantify the total randomness due to measurement sequentially starting from $A_1$, then randomness of $A_2$ given that we know about $A_1$, randomness of $A_3$ given that we know about $A_1$, $A_2$ and so on. The measurement induced randomness (MIR) in this case is given by:

\begin{eqnarray*}
\begin{split}
\Delta_{R}^{A_1A_2...A_N} &= S(\rho_{A_1})+ S(\rho_{A'_2 \mid A'_1}) \\
						  &+ S(\rho_{A'_3\mid A'_1A'_2}) + S(\rho_{A'_4\mid A'_1A'_2A'_3}) \cdots\\
						  &+ S(\rho_{A'_N\mid A'_1A'_2...A'_{N-1}}) - S(\rho_{A_1A_2...A_N}).
\end{split}
\end{eqnarray*}

\section{Interpreting Measurement Induced Randomness: A State Merging Perspective}\label{state_merging}
   
Here in this section we show that the measurement induced randomness (MIR) can be interpreted as the change in the cost of state merging\cite{HOW} for two qubits, three qubits and finally for $N$ qubit systems. 

\subsection{State Merging for two qubit system}
 
In this subsection we show that for two qubit states, the minimum value of the measurement induced randomness (MIR) gives the markup in the cost of quantum state merging due to measurement. To illustrate this, a quantum operation $\mathcal{E}$ is performed on $A_1$ assuming that a unitary $U$ acts on $A_1$ and a pure ancilla state $C$ (initialized to $(\ket{0})$. We note the following observations:
\begin{enumerate}
\item\label{unitary_mutual} $I(A_2:A_1C) = I(A'_{2}:A'_{1}C')$ since unitary interactions do not affect mutual information between subsystems.
\item\label{discard_mutual} $I(A'_{2}:A'_1)\leq I(A'_{2}:A'_1C')$ because dropping out subsystems cannot improve correlations.
\end{enumerate}
The change in the cost of state merging due to measurement is $\Delta = S(\rho_{A'_2|A'_1})- S(\rho_{A_2|A_1})=I(A_{2}:A_{1}) - I(A'_2:A'_1)$. From ~\ref{unitary_mutual} and ~\ref{discard_mutual}, 
we get $I(A_{2}:A_{1}) - I(A'_2:A'_1)\geq I(A_{2}:A_{1}) - I(A'_2 : A'_1C') = I(A_{2}:A_{1}) - I(A_{2} : A_1C) = I(A_{2}:A_{1}) - I(A_{2} : A_1) = 0$. 
Therefore the cost of state merging always increases after measurement. As we shall see, the $\Delta$ is exactly equal to the measurement induced randomness (MIR) for two qubits systems. The state $\rho_{A_1,A_2}$ after measurement reduces to
$\rho_{A_1'A_2'} = \displaystyle\sum_{j}p_{j}\rho_{A_2\mid j}\otimes\pi_j$, where $\pi_j$ are projective measurements. The individual density matrices after measurement are $\rho'_{A_2} = \displaystyle\sum_{j} p_{j}\rho_{A_2\mid j} = \rho_{A_2}$ and $\rho'_{A_1}=\displaystyle\sum_{j}p_{j}\pi_j$. We have
\begin{eqnarray*}
\begin{split}
I(A'_2:A'_1) &= S(\rho_{A'_2}) + S(\rho_{A'_1}) - S(\rho_{A'_1,A'_2}) \\
                 &= S(\rho_{A'_2}) -\sum_{j} p_{j} S(\rho_{A_2\mid\pi_{j}^{A_1}}).                 
\end{split}
\end{eqnarray*}
The increase in the cost of state merging is given by $D =S(\rho_{A'_2|A'_1})- S(\rho_{A_2|A_1})= I(A_{2}:A_{1}) - I(A_{2}':A_{1}')=S(\rho_{A_{2}})+S(\rho_{A_{1}})-S(\rho_{A_1,A_2})-[S(\rho'_{A_{2}})-\displaystyle\sum_jp_jS(\rho_{A_2\mid\pi_{j}A_1})]=S(\rho_{A_1})+S(\rho_{A'_2\mid A'_1})-S(\rho_{A_1A_2})=\Delta_{R}^{A_1A_2}$. This shows that the randomness generated in the system can also be accounted for increase in the cost of state merging. This also shows that for two qubit system there is always increase in the cost of state merging and subsequently measurement induced randomness is a positive quantity for two qubit system.

\subsection{State Merging for three qubit system}

For three qubit systems, we show that total change in the cost of state merging is given by the measurement induced randomness (MIR). We simulate an arbitrary quantum operation $\varepsilon$ (including measurement) on $A_1$. For that, we initially bring in a pure state $D$ ($| 0\rangle$) in proximity to the qubit $A_1$. We assume that $U$ is a unitary interaction between $A_1$ and $D$. Here primes are used denote the state of the systems after $U$ has acted upon. We have $S(\rho_{A_1,A_2,A_3})=S(\rho_{A_1,A_2,A_3,D})$ as $D$ starts with product state with $(A_1,A_2,A_3)$. We also have $I(A_2,A_3:A_1,D)=I(A_2,'A_3':A_1,'D')$ as there is no change in the total correlation of the system as result of unitary interaction. Since discarding quantum systems cannot increase the mutual information, 
$I(A_2',A_3':A_1')\leq I(A_2',A_3':A_1',D')=I(A_2,A_3:A_1,D)=I(A_2,A_3:A_1)$. In other words in terms of conditional entropy we can say at most, $S(\rho_{A_2',A_3'\mid A_1'})\geq S(\rho_{A_2,A_3\mid A_1})$. However, from this we can not conclude whether $S(\rho_{A_2'\mid A_1'})\leq S(\rho_{A_2\mid A_1})$ or $S(\rho_{A_2'\mid A_1'})\geq S(\rho_{A_2\mid A_1})$. The change in the cost of state merging is captured by the quantity $\Delta_1= S(\rho_{A_2'\mid A_1'})-S(\rho_{A_2\mid A_1})$.

The state of $\rho_{A_1,A_2,A_3}$ after measurement on the sub-system $A_1$ changes to $\rho_{A_1',A_2'A_3'} = \displaystyle\sum_j p_j\pi_j^{A_{1}}\otimes\rho_{A_2\mid j}\otimes \rho_{A_3\mid j}$ (where $\pi_j$ is the projection operator). 

Next we consider the case when instead of one particle measurement, we carry out two particle measurement. For that we simulate an arbitrary quantum operation $\varepsilon$ (including measurement) on $A_1$ and $A_2$. We bring in an ancilla $E$ which is initially in a pure state $|0\rangle$. Here $U$ once again is the unitary interaction between $A_1,A_2$ and $E$. Since $E$ is in a complete product state with the rest of the system, it does not contribute to the entropy of the system. We have $S(\rho_{A_1,A_2,A_3})=S(\rho_{A_1,A_2,A_3,E})$. Since unitary interaction does not change the total correlation of the system we have, $I(A_3:A_1,A_2,E)=I(A_3':A_1',A_2',E')$. As we know, by discarding quantum system can not increase the mutual information, we have $I(A_3':A_1',A_2')\leq I(A_3':A_1',A_2',E')=I(A_3:A_1,A_2,E)=I(A_3:A_1,A_2)$. Equivalently, we can write $S(\rho_{A_3'\mid A_1',A_2'})\geq S(\rho_{A_3\mid A_1,A_2})$. Hence, the change in the cost of state merging $\Delta_2= S(\rho_{A_3'\mid A_1',A_2'})-S(\rho_{A_3\mid A_1,A_2})= I(A_3:A_1,A_2) - I(A'_3:A'_1,A'_2) = S(\rho_{A_3}) + S(\rho_{A_1A_2}) -S(\rho_{A_1A_2A_3}) - [S(\rho_{A_3}) -\sum_{j} p_{j} S(\rho_{A_3\mid\pi_{j}^{A_1,A_2}})] = S(\rho_{A_1A_2}) - S(\rho_{A_1A_2A_3}) + S(\rho_{A'_3\mid A'_1,A'_2})$. Similarly, the state of $\rho_{A_1,A_2,A_3}$ after measurement on the subsystem $A_1,A_2$ changes to $\rho_{A_1'A_2'A_3'}=\displaystyle\sum_j p_j\pi_j^{A_1,A_2}\otimes\rho_{A_3\mid j}$ (where $\pi_j$ are two particle projection operators). 

Therefore, the total change in the cost of state merging after carrying out both types  measurement is the total of the change in the cost in each of these individual measurements. This total change in the cost is given by, $\Delta=\Delta_1+\Delta_2= S(\rho_{A_2'\mid A_1'})-S(\rho_{A_2\mid A_1})+ S(\rho_{A'_3\mid A'_1, A'_2})-S(\rho_{A_3\mid A_1,A_2})=S(\rho_{A_1})+S(\rho_{A'_2\mid A'_1})- S(\rho_{A_1A_2}) +  S(\rho_{A_1A_2}) - S(\rho_{A_1A_2A_3}) + S(\rho_{A'_3\mid A'_1,A'_2}) = S(\rho_{A_1}) + S(\rho_{A'_2\mid A'_1}) + S(\rho_{A'_3\mid A'_1,A'_2}) - S(\rho_{A_1A_2A_3})=\Delta_{R}^{A_1A_2A_3}$. However in three qubit case we cannot comment in general that whether the total change in cost as well as the measurement induced randomness is positive or not.\\

\subsection{State Merging for $N$ qubit system}

The same analysis goes through for $N$ qubit systems. We consider measurement performed on $r$ parties where $1\leq r\leq N-1$. Towards this, we simulate an arbitrary quantum operation $\epsilon$ (including measurement) on $A_1,A_2,\cdots,A_r$ by assuming an ancilla $E$ (initially in a pure state $|0\rangle$) and a unitary $U$ between the qubits $A_1,A_2,\cdots,A_r$. We have $S\left(\rho_{A_1},\rho_{A_2},\cdots \rho_{A_n}\right) = S\left(\rho_{A_1},\rho_{A_2},\cdots,\rho_{A_n},E\right)$ as $E$ starts out with a product state with $\left(A_1,A_2,\cdots A_r\right)$ and $I\left(A'_{r+1}\cdots A'_{N}:A'_{1}\cdots A'_r,E'\right)=I\left(A_{r+1}\cdots A_{N}:A_{1}\cdots A_r,E\right)$ because unitary interactions do not affect mutual information. Since discarding quantum systems cannot decrease mutual information, we have
\begin{eqnarray}
&&I\left(A'_{r+1}\cdots A'_{N}:A'_{1}\cdots A'_r\right)  {}\nonumber\\&&
\leq I\left(A'_{r+1}\cdots A'_{N}:A'_{1}\cdots A'_r,E'\right){}\nonumber\\&&
= I\left(A_{r+1}\cdots A_{N}:A_{1}\cdots A_r,E\right) {}\nonumber\\&&
= I\left(A_{r+1}\cdots A_{N}:A_{1}\cdots A_r\right).
\end{eqnarray}

The state of the system after measurement is $\rho_{A'_1,\cdots,A'_N} = \sum_{j}p_{j}\pi_j^{A_{1},\cdots,A_r}\otimes\rho_{A_{r+1}\mid j}\otimes\cdots\otimes\rho_{A_{N}\mid j}$, where $\pi_j$ are $r$ particle projection operators. The net change in the cost of state merging post measurement is given by:
\begin{eqnarray*}
\begin{split}
\Delta &= \sum_{r=1}^{N-1}S\left(\rho_{A'_{r+1}\mid{A'_{1},\cdots,A'_r}}\right) - S\left(\rho_{A_{r+1}\mid{A_{1},\cdots,A_r}}\right) \\
       &= \sum_{r=1}^{N-1}S\left(\rho_{A'_{r+1}\mid{A'_{1},\cdots,A'_r}}\right) + S\left(\rho_{A_{1},\cdots,A_r}\right)  \\
       &-  S\left(\rho_{A_{1},\cdots,A_{r+1}}\right) = S\left(\rho_{A_{1}}\right)- S\left(\rho_{A_{1},\cdots,A_{N}}\right)  \\
       &+ \sum_{r=1}^{N-1}S\left(\rho_{A'_{r+1}\mid{A'_{1},\cdots,A'_r}}\right) \\
       &= \Delta_{U}^{A_1A_2...A_N}.
\end{split}
\end{eqnarray*}
Hence we recover the result that the change in the cost of state merging is exactly equal to the 
measurement induced randomness.

\section{Conclusion}\label{conclusion}
In this work we bring out a very fundamental difference in quantifying the randomness present in a quantum mechanical system. We show that Von neumann entropy alone is not sufficient to quantify the entire randomness present in the system. We introduce a randomness that can be generated in the system as an act of measurement and also find a way to quantify this. We call this as measurement induced randomness (MIR). Further we show that this randomness can also be interpreted as the change in the cost of state merging process as a result of discarding prior information. In short the loss of information in an information processing tasks can also be interpreted as the randomness generated as a result of measurement.

\section{Acknowledgment}
We acknowledge Dr Nilanjana Dutta for useful discussions and thank Mr. Avijit Misra for his help in certain technical areas.


\end{document}